\documentclass[conference]{IEEEtran}
\IEEEoverridecommandlockouts
\usepackage{cite}
\usepackage{amsmath,amssymb,amsfonts}
\usepackage{algorithmic}
\usepackage{graphicx}
\usepackage{textcomp}
\usepackage{xcolor}
\def\BibTeX{{\rm B\kern-.05em{\sc i\kern-.025em b}\kern-.08em
    T\kern-.1667em\lower.7ex\hbox{E}\kern-.125emX}}
\begin{document}

\title{Adversarial attacks on deep learning models for fatty liver disease classification by modification of ultrasound image reconstruction method
}


\author{\IEEEauthorblockN{Micha\l{} Byra\IEEEauthorrefmark{1}\IEEEauthorrefmark{2}, Grzegorz Styczynski\IEEEauthorrefmark{3}, Cezary Szmigielski\IEEEauthorrefmark{3}, Piotr Kalinowski\IEEEauthorrefmark{4}, Lukasz Michalowski\IEEEauthorrefmark{5}, \\ Rafal Paluszkiewicz\IEEEauthorrefmark{4}, Bogna Ziarkiewicz-Wroblewska\IEEEauthorrefmark{5}, Krzysztof Zieniewicz\IEEEauthorrefmark{4}, Andrzej Nowicki\IEEEauthorrefmark{2} }

\IEEEauthorblockA{\IEEEauthorrefmark{2}Department of Ultrasound, Institute of Fundamental Technological Research, \\Polish Academy of Sciences, Warsaw, Poland}
\IEEEauthorblockA{\IEEEauthorrefmark{3}Department of Internal Medicine, Hypertension and Vascular Diseases, Medical University of Warsaw, Poland}
\IEEEauthorblockA{\IEEEauthorrefmark{4}Department of General, Transplant and Liver Surgery, Medical University of Warsaw, Poland}
\IEEEauthorblockA{\IEEEauthorrefmark{5}Department of Pathology, Center for Biostructure Research, Medical University of Warsaw, Poland}
\IEEEauthorblockA{\IEEEauthorrefmark{1}Corresponding author, e-mail: byra.michal@gmail.com}
}

\maketitle

\begin{abstract}

Convolutional neural networks (CNNs) have achieved remarkable success in medical image analysis tasks. In ultrasound (US) imaging, CNNs have been applied to object classification, image reconstruction and tissue characterization. However, CNNs can be vulnerable to adversarial attacks, even small perturbations applied to input data may significantly affect model performance and result in wrong output. In this work, we devise a novel adversarial attack, specific to ultrasound (US) imaging. US images are reconstructed based on radio-frequency signals. Since the appearance of US images depends on the applied image reconstruction method, we explore the possibility of fooling deep learning model by perturbing US B-mode image reconstruction method. We apply zeroth order optimization to find small perturbations of image reconstruction parameters, related to attenuation compensation and amplitude compression, which can result in wrong output. We illustrate our approach using a deep learning model developed for fatty liver disease diagnosis, where the proposed adversarial attack achieved success rate of 48\%.

\end{abstract}


\section{Introduction}

Convolutional neural networks (CNNs) have achieved remarkable success in medical image analysis tasks, such as image classification, object detection and semantic segmentation. However, CNNs can be vulnerable to adversarial attacks, even small perturbations applied to input data may significantly affect model performance and result in wrong output \cite{goodfellow2014explaining,akhtar2018threat,qiu2019review}. Such perturbations may occur accidentally or can be intentionally designed with the aim to fool the model, for example by direct modification of the input image pixel intensities. Existence of adversarial examples raises questions about the robustness of deep learning models, and is especially important in medical imaging. In ultrasound (US) imaging, the appearance of tissues depends on the applied image reconstruction method. US image pixel intensities depend on the attenuation compensation technique and the algorithms used to process radio-frequency (RF) US signals. Nonlinear compression of US echoes may enhance the visibility of tissue interfaces, but also remove speckle patterns specific to particular tissues. In practice, radiologists and physicians use different scanner settings to obtain the desired US image quality. However, as presented in the previous studies, modifications of US image pixel intensities may affect extraction of texture features and lower the performance of US based machine learning models \cite{byra2019quantitative, byra2019impact,gomez2020assessment}. In the case of other medical imaging modalities, the vulnerability of deep learning models to adversarial attacks has been presented, among others, using dermoscopy images \cite{finlayson2018adversarial}. 

In this work, we devise a novel approach to adversarial attacks, which is specific to US imaging. In comparison to methods from computer vision that directly modify image pixel intensities, we explore the possibility of fooling deep learning model by perturbing B-mode US image reconstruction method \cite{akhtar2018threat}. We investigate whether a modification of the reconstruction method may change the distribution of US image pixel intensities, and consequently result in wrong output. We illustrate the proposed approach using a deep learning model developed for the diagnosis of fatty liver disease, which is an important medical problem \cite{beeman2018imaging,wong2018noninvasive}. First, we use transfer learning to develop a deep learning model for classification of liver US images. Second, we apply zeroth order optimization (ZOO) to find a perturbation in the space of image reconstruction parameters, which results in wrong classification of the inputted reconstructed US image. To the best of our knowledge, while several groups have proposed deep learning models for fatty liver disease diagnosis, the robustness of such methods to adversarial attacks has not been investigated yet \cite{byra2018transfer,han2020noninvasive,cao2020application}. 

\section{Methods}

\subsection{Dataset}

To develop the deep learning model and to assess the proposed adversarial attack, we used the following datasets: 

\begin{enumerate}
    \item 178 US images collected from 178 patients with the GE Vivid E9 System (GE Healthcare INC, Horten, Norway).
    \item 33 RF data frames (post-beamformed, before US image reconstruction) collected from 33 patients with Siemens System (Siemens, Issaquah, Wash). 
\end{enumerate}

The data were collected from the liver/kidney view from patients admitted for bariatric surgery. The scanning was performed with convex transducers operating at imaging frequency of around 2.5 MHz. Fatty liver disease was diagnosed based on liver biopsy (more than 5\% hepatocytes with steatosis). For  each  dataset, approximately 65\% of patients had fatty liver disease. Several US images from each dataset are presented in Fig. 1.

\subsection{Image reconstruction}

Commonly, the reconstruction method includes several procedures, e.g. attenuation compensation, compression of RF signals, data interpolation and resizing. In our case, the reconstruction of US images based on RF signals included the following steps:

\begin{enumerate}
    \item Amplitude calculations with the Hilbert transform.
    \item Attenuation compensation based on fixed attenuation coefficient $\beta$.
    \item Interpolation from the coordinate space of the convex transducer to correct spatial dimensions. 
    \item Logarithmic compression of amplitude samples and thresholding to specific decibel range specified by upper $\alpha_u$ and lower $\alpha_l$ threshold levels (e.g. amplitude samples below $\alpha_l$ were set to $\alpha_l$). 
    \item Mapping of compressed and thresholded amplitude samples to US image pixel intensities (8 bits). 
\end{enumerate}

In this work, we performed the adversarial attack based on the modification of the attenuation coefficient $\beta$ and the compression threshold levels $\alpha_u$ and  $\alpha_l$. We selected these parameters, because any modification of these parameters directly affects US image pixel intensities, and consequently change the appearance of edges and speckle patterns in US image. For the sake of the experiments, we also selected the following initial reconstruction parameters: $\beta$=0.9 dB/(cm*MHz), $\alpha_{l}$=10 dB, $\alpha_{u}$=55 dB. These parameters were used as a starting point for the search of the adversarial perturbation. Selection of the attenuation coefficient was motivated by the previous studies on fatty liver disease assessment with quantitative US \cite{han2019inter,han2020assessment}. The compression threshold levels were selected based on subjective visual assessment of differently reconstructed US images. 

\begin{figure}[t]
	\begin{center}
		\includegraphics[width=1\linewidth]{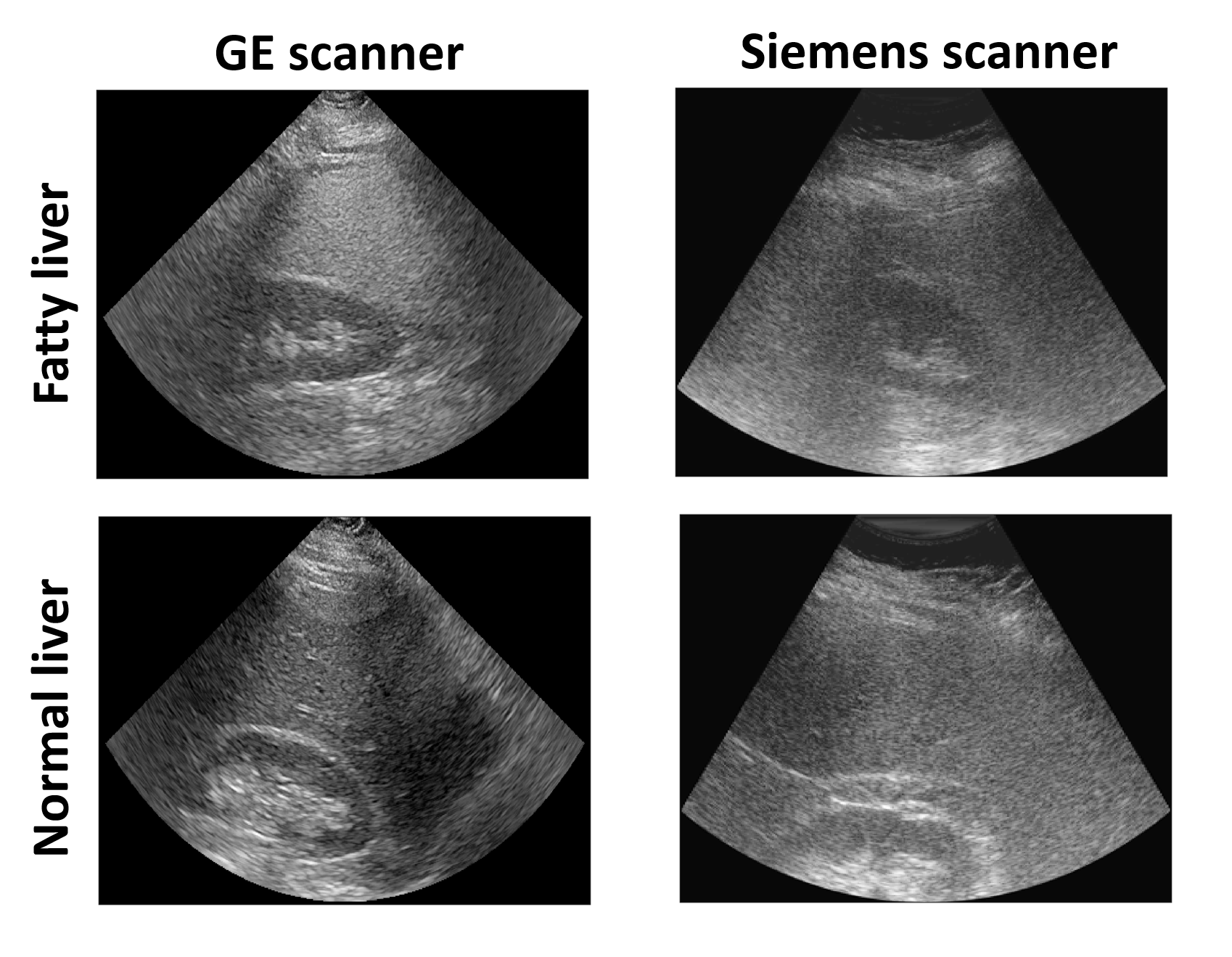}
	\end{center}
	\caption{US images acquired with the two scanners and used in our study to investigate the proposed adversarial attack.}
	\label{pipe}
\end{figure}

\begin{figure*}[t]
	\begin{center}
		\includegraphics[width=1\linewidth]{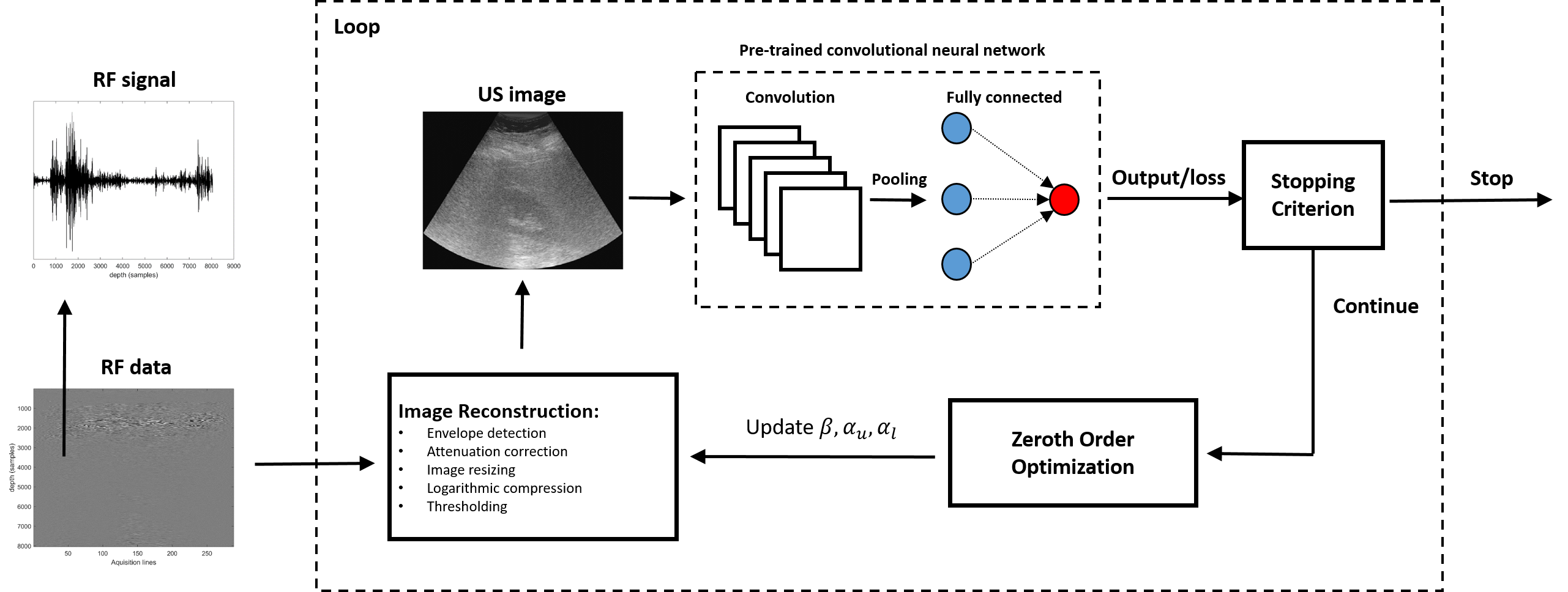}
	\end{center}
	\caption{Pipeline of the investigated adversarial attack on US based deep learning model. The reconstruction parameters $\beta, \alpha_{l}, \alpha_{u}$ are updated using zeroth order optimization to find a set parameters resulting in wrong classification. }
	\label{pipe}
\end{figure*}

\begin{figure}[]
	\begin{center}
		\includegraphics[width=1\linewidth]{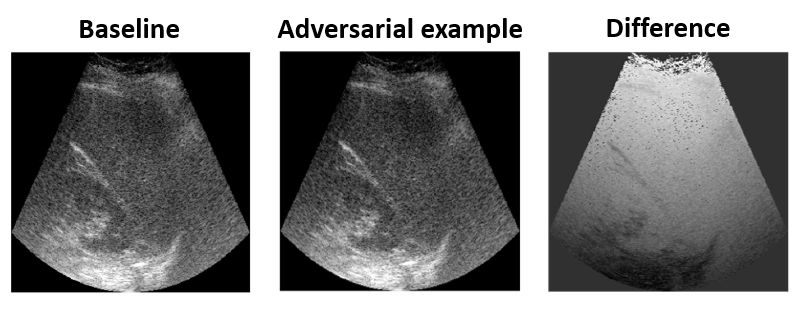} 
	\end{center}
	\caption{US image presenting fatty liver (reconstruction parameters $\beta$=0.9 dB/(cm*MHz), $\alpha_{l}$=10 dB, $\alpha_{u}$=55 dB) and the corresponding adversarial example (reconstruction parameters $\beta$=1.05 dB/(cm*MHz), $\alpha_{l}$=11.5 dB, $\alpha_{u}$=54.5 dB, and the difference between the two images. }
	\label{pipe}
\end{figure}

\subsection{Deep learning model}

We used the InceptionResNetV2 CNN pre-trained on the ImageNet dataset for the fatty liver disease diagnosis \cite{russakovsky2015imagenet,szegedy2017inception}. The last dense layer was replaced with a randomly initialized dense layer suitable for the binary classification. The network was trained using the first dataset and with the images from the second dataset reconstructed based on the initial parameters. Stochastic gradient descent algorithm was applied to minimize the binary cross-entropy loss. Dropout regularization and early stopping were applied to address the over-fitting problem. Loss function was weighted to take into account class imbalance. Moreover, the reconstructed gray-scale US images were resized to (299, 299), the original input size the InceptionResNetV2 CNN, and duplicated across all color channels to imitate RGB images. These two steps may also be considered as the part of the US input image reconstruction method. 

\subsection{Adversarial attack}

The proposed method works in a black-box setting, it only requires the access to the output of the network to calculate the loss function (binary cross-entropy in our case) \cite{chen2017zoo,liu2018signsgd}. The attack is performed for each RF data frame separately. We reconstruct the first US image using the initial parameters, and next for each step of the procedure we determine a perturbation resulting in performance decrease. While there is no explicit formula to calculate the gradient of the loss function in respect to reconstruction parameters, we can approximate the gradient of the, for example, attenuation coefficient $\beta$ with the following formula: 

\begin{equation}
    \frac{\partial J}{\partial \beta} \approx \frac{ J(\beta+\delta \beta) - J(\beta-\delta \beta)}{2\delta \beta},
\end{equation}

\noindent where J($\cdot$) stands for the binary cross-entropy loss depending on the input image, image label, network parameters, reconstruction parameters $\beta, \alpha_{l}, \alpha_{u}$ and $\delta \beta$ is the step size set for the $\beta$ parameter. Similarly, this formula can be used to calculate the gradient in respect to the $\alpha_{l}$ and $\alpha_{u}$ parameters. Given the gradients, we can apply sign coordinate gradient descent and update the reconstruction parameters in a way to maximize the loss function and undermine the model. For example, to update the $\beta$ parameter we can apply the following formula:

\begin{equation}
    \beta_{i+1} = \beta{i} + \epsilon_{\beta} \text{sign} \left(\frac{\partial J}{\partial \beta} \right),
\end{equation}

\noindent where $i$ stands for the $i$-th iteration step of the procedure and $\epsilon_{\beta}$ is the learning rate for the $\beta$ parameter. The remaining reconstruction parameters, $\alpha_{l}$ and $\alpha_{u}$, can be updated in a similar way. 

In the study, we determined the steps in eq. 1 and the learning rates in eq. 2 experimentally. We found that steps equal to 0.05, 0.1, 0.1 and the learning rates equal to 0.05, 0.5, 0.5 for the $\beta, \alpha_{l}, \alpha_{u}$ performed well in our case. The pipeline of the proposed adversarial attack is illustrated in Fig. 2. To assure relatively small perturbations, we limited the min/max ranges of the possible reconstruction parameters to (0.5, 1.3), (5, 15) and (50, 60) for the $\beta, \alpha_{l}$ and $\alpha_{u}$ parameter, respectively. Procedure was stopped after obtaining perturbation resulting in wrong classification or after reaching the min/max parameters. The cut-off for the classification was set to 0.5. All computations performed in this work were done in Python, the deep learning model was implemented in TensorFlow \cite{abadi2016tensorflow}. 

\section{Results}

The area under receiver operating characteristic curve (AUC) and accuracy in the case of the second dataset were equal to 0.84 and 0.82 (27/33), respectively. The correctly classified cases were utilized to assess the proposed adversarial attack. We were able to perform successful attack (resulting in misclassification) on 13 out of 27 cases, with success rate of 48\%. We found that all reconstruction parameters were used by the optimizer to minimize the loss and perturb the data. In the case of the unsuccessful attacks, the procedure reached the parameter bounds each time, but the change in the network output was to small to result in misclassification. Fig. 3 presents an example of a successful adversarial attack
on an US image presenting fatty liver.    

\section{Discussion}

In this work, we presented that adversarial attacks based on the modification of US image reconstruction method are feasible. Our approach was demonstrated with a deep model developed for fatty liver disease diagnosis. As presented in Fig. 3, even small change of the parameters related to the reconstruction method may significantly decrease classification performance of the deep model and result in wrong output. This might be probably due to complex behaviors of deep models \cite{goodfellow2014explaining,akhtar2018threat}. In comparison to the adversarial attacks from computer vision, which aim to directly modify input image pixels, our approach is specific to US imaging. In our case, the modification of the US image pixels results from the change of the image reconstruction method. Nevertheless, while we designed the perturbations, accidental changes of the reconstruction method may also arise in practice due to, for example, modification of the scanner settings. The network utilized in our study was trained with the US images from the second dataset reconstructed using the same initial parameters. Taking this into account, our study suggests that to develop more efficient US based deep learning models it might be necessary to augment training data with differently reconstructed US images. Our study also suggests that the RF data might serve as a better data type for training of the deep models, because training based on RF data does not require US image reconstruction and therefore should be more robust.   

In future, we would like to apply the ZOO technique to study the robustness of different machine learning models. The proposed approach to robustness assessment is general, it can be applied to examine machine learning models based on handcrafted texture features and standard classifiers (e.g. support vector machines, random forests). We would like also to expand our approach by taking into account other parameters related to US image reconstruction, for example those related to image scaling and filtration. In this work, we applied a relatively simple optimization method to find the adversarial perturbations, but in future it would be interesting to also investigate other methods. Moreover, we did not examine potential defences to the attack. For example, the augmentation of the training set with differently reconstructed images would probably result in better performance and a deep model that would be more robust to adversarial attacks. 

\section*{Conflict of interest}

The authors do not have any conflicts of interest. 

\bibliographystyle{IEEEtran}
\bibliography{IEEEabrv,references}

\end{document}